\documentclass{article}
\usepackage[latin1]{inputenc}
\usepackage{url}
\usepackage{graphicx}

\begin{document}

\title{Who is the best connected EC researcher? Centrality analysis of the complex network of authors in evolutionary computation}

\author{Juan-J. Merelo and Carlos
  Cotta\thanks{JJ Merelo ({\sf jj@merelo.net}) is with the Depto. ATC,
    U. Granada, and Carlos Cotta ({\sf ccottap@lcc.uma.es}) with the
    Depto. LCC, U. Malaga, Spain}} 

\maketitle
\begin{abstract}
Co-authorship graphs (that is, the graph of authors linked by
co-authorship of papers) are complex networks, which expresses the
dynamics of a complex system. Only recently its study has started to
draw interest from the EC community, the first paper dealing with it
having been published two years ago. In this paper we will study the
co-authorship network of EC at a microscopic level. Our objective is
ascertaining which are the most relevant nodes (i.e. authors) in it.
For this purpose, we examine several metrics defined in the
complex-network literature, and analyze them both in isolation and
combined within a Pareto-dominance approach. The result of our
analysis indicates that there are some well-known researchers that
appear systematically in top rankings. This also provides some hints
on the social behavior of our community.

{\bf Keywords}: Social networks, co-authorship networks, scientometrics,
  sociology of science, evolutionary computation, eigenvalues, Pareto
  front 
\end{abstract}

\section{Introduction}
\label{sec:intro}

Academy, as any other human endeavor, is a complex adaptive system (CAS), and looking at some of
its aspects reflect that fact. Since one of the outstanding (and
measurable) aspects of academy is publishing, it is interesting to
study it to find out how general CAS mechanisms apply to it, and
create general models of the system.

One of the possible ways to study this publishing activity is to
look at co-authorship graphs, where nodes (or {\em actors}) correspond to
authors, joined by an {\em edge} if they have been coauthors in a
paper. This is a non-directed graph, which does not take into
account the authorship order, and, besides, considers that all
signing authors are {\em actually} authors: following Yoshikane
\emph{et al.}~\cite{yoshikane06}, we consider that, in general, this
assumption is true. Co-authorship graphs have been studied for a long
time, starting with Kretschmer~\cite{kretschmer94}, but they started
to be recognized as complex networks with the work of
Newman~\cite{newman01a,newman01b} and Barab\'asi \cite{barabasi02},
showing they followed power-laws \cite{measur:jeong2003} (which
might correspond to a preferential attachment growth) and also
behaved as a small world \cite{measur:jeong2003}.

Even as the general framework has been already laid out, there are
still a few open issues. Measurements for a particular field, such as
evolutionary computation \cite{ec-network-2006}, have to be made, and
the evolution of its graph followed~\cite{cotta07:gpem}. This
evolution reflects the differential authoring
mechanisms in particular fields, and these mechanisms can be
modelled. Besides, within every field, finding out sociometric stars
reflect the knowledge flow within it and its fertility. Synchronic
networking (co-authorship relations) are also related to diachronic
networking (citing or co-citing relations), and, thus, it is also
interesting for knowledge discovery within a particular field.

Another open issue is exactly what to measure in that network.
Looking at a single measure will yield a partial view of the
network. While there is a high correlation among some measures (such
as betweenness and closeness; definitions will follow later), they
reflect different aspects of the network and, thus, they will have to
be taken into account globally when making a ranking of the
sociometric stars of the graph. In this paper, each actor will be
assigned a vector of quantities, and the ranking will be done
according to the concept of {\em Pareto front} \cite{horn:1994:npamo},
that is, the set of non-dominated authors. Identifying the key actors
in our field not only provides some objective metric with which our
subjective perception can be contrasted, but it can be also helpful in
order to understand some of the patterns of social behavior at work in
our community.

The rest of the paper is organized as follows: next we present a
brief state of the art on the subject. The resources and methodology
used in this paper are presented in section \ref{sec:met}, and the
results of the analysis in section \ref{sec:stars}. Finally, we will
draw some conclusions in section \ref{sec:conclusions}.

\section{State of the art}
\label{sec:soa}

Coauthorship studies have generally focused in macroscopic measures of
particular scientific communities:
computer support of cooperative work \cite{horn2004}, psychology
and philosophy \cite{psy2003}, chemistry \cite{chem2004}, SIGMOD
authors \cite{SIGMOD2003} and sociology \cite{moody04}, but some other
authors \cite{Cardillo06} have analyzed the topological properties
of these networks in general, looking at a particular preprints
database ({\sf cond-mat}). They have found that betweenness and
degree follow a long-tailed degree distribution, which is usually
to be expected, but it is interesting to prove that it actually
happens for a representative sample. 

Although all coauthors are considered indistinctly, insome cases the roles of different authors
\cite{yoshikane06} is taken into account, although their focus is on
visualization of relations among authors, not on a differential
analysis of the different positions, or on the study of the internal
structure of the cliques formed via co-authorship. 

Another approach to the study of scientific communities is to
understand the role of different actors by measuring certain
microscopic (node-based) quantities; centrality (see
\cite{borgatti06:centrality}) is one of them, although its definition
is not trivial. For
example: is an actor more central when it is more visible, or more
influential, or more powerful? Furthermore, how can
visibility, influence, and power be defined in an objective sense?
It should be taken into account that intuition can be misleading
here. As an example, consider 
that one of the first (intuitive) principles that were proposed is that
centrality grows monotonically with the number of ties, and that
adding ties (edges) should increase one node's centrality
\cite{Sabidussi66}. While these ideas look attractive and intuitive,
they do not provide a satisfactory definition of centrality, since
the importance of a certain node can however be diminished when
other node gets more ties. Freeman \cite{freeman79} gave an answer
to this issue, reviewing a number of published measures, and
identified three basic concepts for defining centrality: degree,
closeness, and betweenness. In this canonical formulation, these
three measures have maximum values when the network is star-shaped,
hence providing a proper characterization of centrality. Borgatti
\cite{Borgatti05} has also elaborated on this issue, considering the
dynamic flow all over the network, and how often traffic flows
through a certain node, or how long does it take to get to a certain
node.

The amount of papers on this area indicates that there is a lot of
work to be done, be it for a particular area or discipline, or on
the visualization and methods front. In this paper, we try to use a
new method -based on Pareto dominance- to examine the sociometric
ranking in a particular field taking into account several centrality
features at the same time.

\section{Resources and methodology}
\label{sec:met}

The bibliographical data used for the construction of the
scientific-collaboration network in EC has been gathered from the
DBLP --\emph{Digital Bibliography \& Library Project}-- bibliography
server, maintained by Michael Ley at the University of Trier. This
database provides bibliographic information on major computer
science journals and proceedings, comprising more than 830,000
articles and several thousand computer scientists (by the end of
2006). We have defined a collection of terms that include the
acronyms of EC-specific conferences --such as GECCO, PPSN or
EuroGP-- or keywords --such as ``Evolutionary Computation'',
``Genetic Programming'', etc.-- that are sought in the title or in
the publication forum of papers. Using an initial sample of authors
(those that have published at least one paper in the last five years
in any of the following large EC conferences: GECCO, PPSN, EuroGP,
EvoCOP, and EvoWorkshops), their list of publications is checked for
relevance, and the corresponding co-authors are recursively
examined. Just as an indication of the breadth of the search, the
number of authors used as seed is 3,773 whereas the final number of
authors in the network is 7,712, that is, more than twice as many.

The macroscopic measures of the  network obtained through this procedure are shown in
\ref{tab:todo}, comparing them with measures taken from a CS
repository (NCSTRL, data taken from \cite{newman04}, and historical
data on what is basically the same network, taken from
\cite{CM05:arxiv,ec-network-2006}. 
\begin{table}[t!]
\begin{center}
\caption[]{Summary of results of the analysis of a computer science
  collaboration network (NCSTRL), the previous analysis of the EC
  co-authorship network --EC05--, taken from
  \cite{ec-network-2006}. Measures for this paper are shown in the middle
  column, and were taken during November 2006. Data for NCSTRL is taken
  from Newman \cite{newman04}. \label{tab:todo} }
\begin{tabular}{lccc}
\hline
                            & EC05    & EC06    & NCSTRL \\
                            \cline{2-4}
total papers                & 6199    & 8501    & 13169  \\
total authors               & 5492    & 7712    & 11994  \\
mean papers per author      & 2.9     & 2.87    & 2.6    \\
mean authors per paper      & 2.56    & 2.60    & 2.22   \\
collaborators per author    & 4.2     & 4.02    & 3.6    \\
size of the giant component & 3686    & 4804    & 6396   \\
\ \ \ \ \ as a percentage   & 67.1\%  & 62.3\%  & 57.2\% \\
2nd largest component       & 36      & 106     & 42     \\
clustering coefficient      & 0.798   & 0.811   & 0.496  \\
mean distance               & 6.1     & 10.9    & 9.7    \\
diameter                    & 18      & 21      & 31     \\
\hline
\end{tabular}
\end{center}
\end{table}

The first and obvious observation is that there has been a
progression from 2005 to 2006: more than two thousand new authors,
but it is also interesting to see that many of these authors have
gone to the main component. This increase in the number of authors
probably account for the increase in the diameter, that goes from 18
to 21, a small increase which shows again the small world
characteristic of this network. Metrics such as the clustering
coefficient, the average number of collaborators per author or
authors per paper are quite close, with a very small variation,
which indicates that collaboration patterns continue in the same
way. We will analyze in next section the microscopic features of the
network, and in particular who the most prominent nodes are.

\section{Sociometric stars}
\label{sec:stars}

Centrality can be measured in multiple ways. We are going to focus
firstly on metrics based on geodesics, i.e., the shortest paths
between actors in the network. These geodesics constitute a very
interesting source of information: the shortest path between two
actors defines a ``referral chain'' of intermediate scientists
through whom contact may be established -- cf. \cite{newman04}. It
also provides a sequence of research topics (recall that common
interests exist between adjacent links of this chain, as defined
by the co-authored papers) that may suggest future joint works or
lines of research.

The first geodesic-based centrality measure we are going to analyze
is \emph{betweenness} \cite{freeman77}, i.e., the relative number of
geodesics between any two actors $j,k$ ($\#g_{jk}$) passing through
a certain $i$ ($\#g_{jik}$), summed for all $j,k$:
\begin{equation}
C^{BET}_i = \sum_j \sum_k \frac{\#g_{jik}}{\#g_{jk}}\ .
\end{equation}

This measure is based on the information flow between actors: when a
joint paper is written, the authors exchange lots of information
(such as knowledge of certain techniques, research ideas, potential
development lines, or unpublished results) which can in turn be
transmitted (at least to some extent) to their colleagues in other
papers, and so on. Hence, actors with high betweenness are in some
sense ``hubs'' that control this information flow; they are
recipients --and emitters-- of huge amounts of cutting-edge
knowledge; furthermore, their removal from the network results in
the increase of geodesic distances among a large number of actors
\cite{wassermanFaust94}.

\begin{table}[h!]
\begin{center}
\caption[]{Top ten actors according to their betweenness.
\label{tab:betweenness} }
\begin{tabular}{rclcr}
\hline
&~~~& Name           &~~~& betweenness \\
\hline
 1 && D.E. Goldberg  && 2194962 \\
 2 && K. Deb         && 1861389 \\
 3 && M. Schoenauer  && 1479185 \\
 4 && H. de Garis    && 1246007 \\
 5 && Z. Michalewicz && 1144581 \\
 6 && X. Yao         && 1060389 \\
 7 && R.E. Smith     &&  928108 \\
 8 && M. Tomassini   &&  921023 \\
 9 && T. B\"ack       &&  818897 \\
10 && K.A. De Jong   &&  772788 \\
\hline
\end{tabular}
\end{center}
\end{table}

\begin{figure}[thbp]
\centering
\includegraphics[width=80mm]{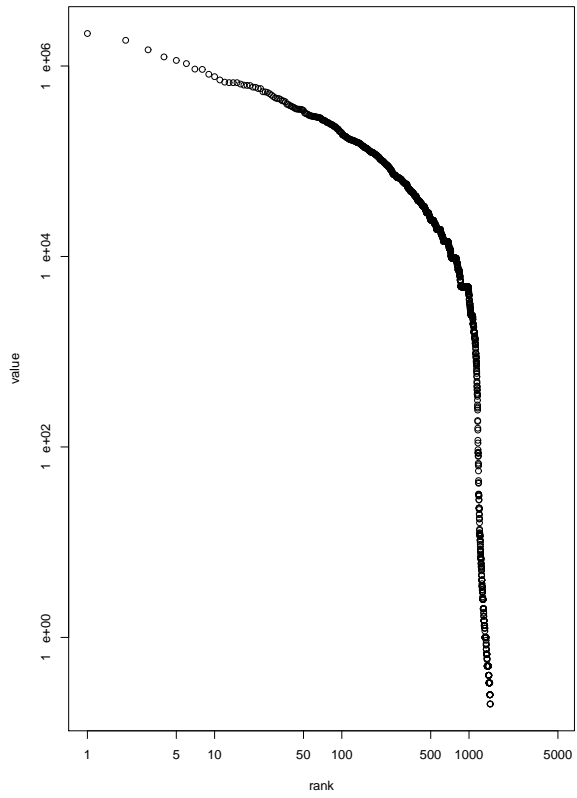}
\caption[]{Logarithmic plot of betweenness vs. rank for the EC
  co-authorship network. There is a power law up to position \#500, with
  an exponential cutoff after that. It is quite usual for social
  networks to have a power law in these quantities.\label{fig:between}}
\end{figure}

Table \ref{tab:betweenness} shows the top ten actors according to
this centrality measure. Notice how the betweenness values decrease
abruptly from one actor to the next. There is clearly a power law at
work (actually, a power law with exponential cutoff), as is shown in
Figure \ref{fig:between}.
This scaling is consistent with the presence of a hierarchy of
hubs in the network. Whenever a shortest path is sought between
two nodes, the nearest common ancestor in the hierarchy is used.
Top actors in this ranking are thus those located in the center of
gravity of the network, connecting distant regions of the latter.

The second centrality measure we are going to consider is
precisely based on this geodesic distance. Intuitively, the length
of a shortest path indicates the number of steps that research
ideas (and in general, all kind of memes) require to jump from one
actor to another. Hence, scientists whose average distance to
other scientists is small are likely to be the first to learn new
information, and information originating with them will reach
others quicker than information originating with other sources.
Average distance (i.e., \emph{closeness}) is thus a measure of
centrality of an actor in terms of their access to information.

\begin{table}[h!]
\begin{center}
\caption[]{Top ten actors according to their closeness.
\label{tab:closeness} }
\begin{tabular}{rclcr}
\hline
&~~~& Name           &~~~& closeness \\
\hline
 1 &&K. Deb      &&4.1886571e-005 \\
2 &&Z. Michalewicz && 4.1091387e-005 \\
3 &&D.E. Goldberg  && 4.0731538e-005 \\
4 &&M. Schoenauer  && 4.0731538e-005 \\
5 &&B. Paechter &&3.9521005e-005 \\
6 &&A.E. Eiben  &&3.9502271e-005 \\
7 &&D.B. Fogel  &&3.8568343e-005 \\
8 &&H.-G. Beyer &&3.8433452e-005 \\
9 &&H.A. Abbass   &&  3.8358266e-005 \\
10&& M. Tomassini   && 3.8165026e-005 \\
\hline
\end{tabular}
\label{tab:centrality}
\end{center}
\end{table}

Table \ref{tab:betweenness} shows the top ten actors according to
this centrality measure, expressed here as the reciprocal of
\emph{farness}, that is, the sum of the lengths of geodesic paths
($d_{ij}$) from a node to every other one:
\begin{equation}
C^{CLO}_i = \frac{1}{\sum_j d_{ij}}\ .
\end{equation}

Notice that the differences in closeness values are not so marked
as for betweenness, i.e. there is no power law acting here. The
fact that our network is a small world contributes to this. Notice
also that the names appearing in this ranking are very similar,
although not identical, to the ranking yielded by betweenness;
there are actually five well-known researchers (K. Deb, D.E.
Goldberg, Z. Michalewicz, M. Schoenauer, and M. Tomassini) showing
up in both rankings, in slightly different places.

In addition to the centrality measures based on geodesics, there
exist an interesting group of metrics based on degree, that is on
the number of ties each actor has. One of them is Bonacich's index,
also called {\em power}~\cite{bonacich87,borgatti06:centrality}. In
social networks, this power is related to the possibility of going
(originally {\em negotiating}) from one to another actor in the network
using all possible paths. Lots of paths imply lots of options,
which, in turn, imply many different ways of negotiating with or
influencing another actor in the network. Notice that this index can
be also interpreted in the following way: one's power is higher is
one has many connections, and even more if these connections have
high power too. Actually, one of the methods for determining this
index is finding the fixed point of a linear combination of one's
degree and one's neighbors' power. The coefficients of this linear
combination have to be chosen so that the procedure converges (this
can be ensured by picking a value lower than the dominant eigenvalue
of the adjacency matrix \cite{borgatti06:centrality}), and the
resulting values are adequately normalized (in our case, the average
squared power is 1.0):
\begin{equation}
C^{POW}_i = \sum_j A_{ij}\left(\alpha+\beta C^{POW}_j\right)\ .
\end{equation}
where $A$ is the adjacency matrix. Results for this measure,
presented for the first time for the EC network in this paper, are
shown in table \ref{tab:power}.

\begin{table}[h!]
\begin{center}
\caption{Top ten actors according to their power (Bonacich power)}
\begin{tabular}{rclcr}
\hline
&~~~& Name           &~~~& power \\
\hline
 1 && D.E. Goldberg && 1.43818\\
 2 && M. Schoenauer && 1.29379\\
 3 && K. Deb        && 1.18711\\
 4 && D. Keymeulen  && 1.05472\\
 5 && X. Yao        && 1.04514\\
 6 && L.D. Whitley  && 1.04259\\
 7 && T. B\"ack      && 0.93871\\
 8 && T. Higuchi    && 0.93698\\
 9 && H. de Garis   && 0.91823\\
10 && L. Kang       && 0.89615\\
\hline
\end{tabular}
\label{tab:power}
\end{center}
\end{table}
There are many names in common with the other rankings, but the
interesting part is precisely those that are not in common, specially
D. Keymeulen. D. Keymeulen works in evolvable hardware, and is a
coauthor with T. Higuchi, who is also a new name in this ranking; in
turn, Dr. Higuchi is coauthor with H. de Garis and X. Yao, which
accounts for the high values of this measure for them. H. de Garis has
also coauthored with L. Kang, explaining out the rest of the ranking
(it must be added that L. Kang is coauthor of Z.  Michalewicz, who is
the 11th in the ranking). Besides, L. Kang and T. Higuchi are {\em
doors} to huge national communities (Chinese and Japanese), over which
they have {\em power}. It is not surprising to find the actors in this
ranking separated by just a few {\em handshakes}, or edges. In fact,
there are several {\em high-ranking kernels}, with each sociometric
figure being a {\em door} or {\em center} to a few of them.

A related approach to the previous one follows when $\alpha=0$. In
that case, an actor's power is simply
\begin{equation}
C^{EIG}_i = \beta\sum_j A_{ij}C^{EIG}_j
\end{equation}
that is, the vector of centrality values is an eigenvector of the
adjacency matrix, and $\beta$ must be the reciprocal of an
eigenvalue. It is customary to pick the centrality vector associated
with the largest eigenvalue. In this case, the resulting ranking is
shown in Table \ref{tab:eigen}.

\begin{table}[h!]
\begin{center}
\caption{Top ten actors according to their eigenvector centrality}
\begin{tabular}{rclcr}
\hline
&~~~& Name           &~~~& score \\
\hline
 1 && D. Keymeulen && 0.358\\
 2 && T. Higuchi   && 0.312\\
 3 && M. Iwata     && 0.255\\
 4 && I. Kajitani  && 0.249\\
 5 && N. Kajihara  && 0.221\\
 6 && M. Murakawa  && 0.205\\
 7 && E. Takahashi && 0.169\\
 8 && H. Sakanashi && 0.166\\
 9 && N. Otsu      && 0.154\\
10 && M. Salami    && 0.153\\
\hline
\end{tabular}
\label{tab:eigen}
\end{center}
\end{table}
As it can be seen, the dominant eigenvector gravitates around D.
Keymeulen, who is the actor with 6th-highest degree (50 coauthors).
His role is enhanced by his collaboration con T. Higuchi (and vice
versa). The remaining actors in the top ten happen to be coauthors
of these two researchers, and hence their presence can be partly
attributed to ``hitchhiking''.

It is evident from the inspection of the above tables that each of
the measures tells a different part of the story. Although there is
some correlation among them, this correlation is not perfect, and
hence the rankings differ. It makes then sense to think about
possible ways of combining this information. This is a standard
issue in multi-objective optimization, and we can think in the
notion of Pareto-dominance as a means to achieve a global
perspective on the centrality status of the different actors. This
is what we do next.

\begin{figure}[htbp]
\centering
\includegraphics[width=85mm]{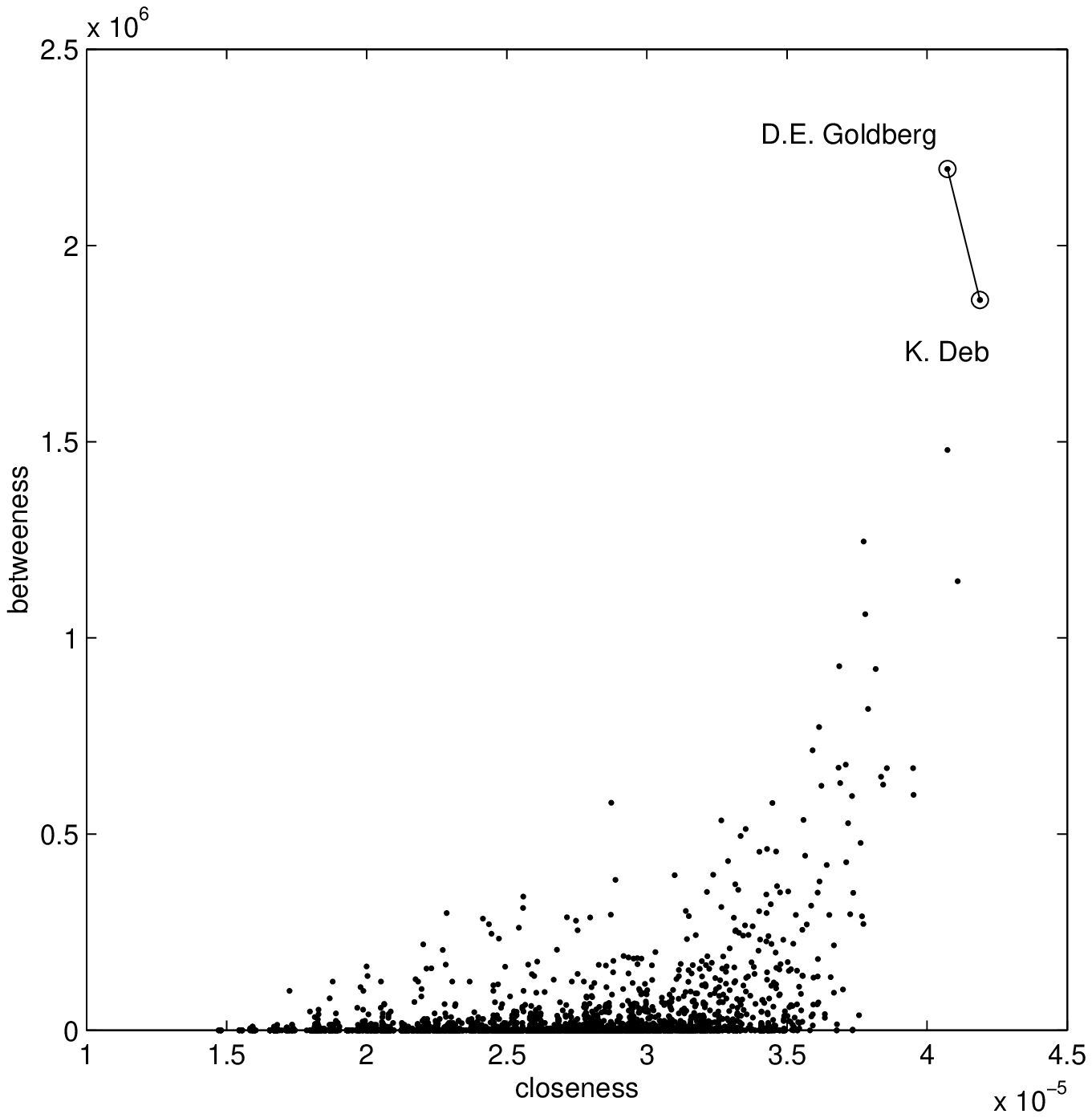}
\includegraphics[width=85mm]{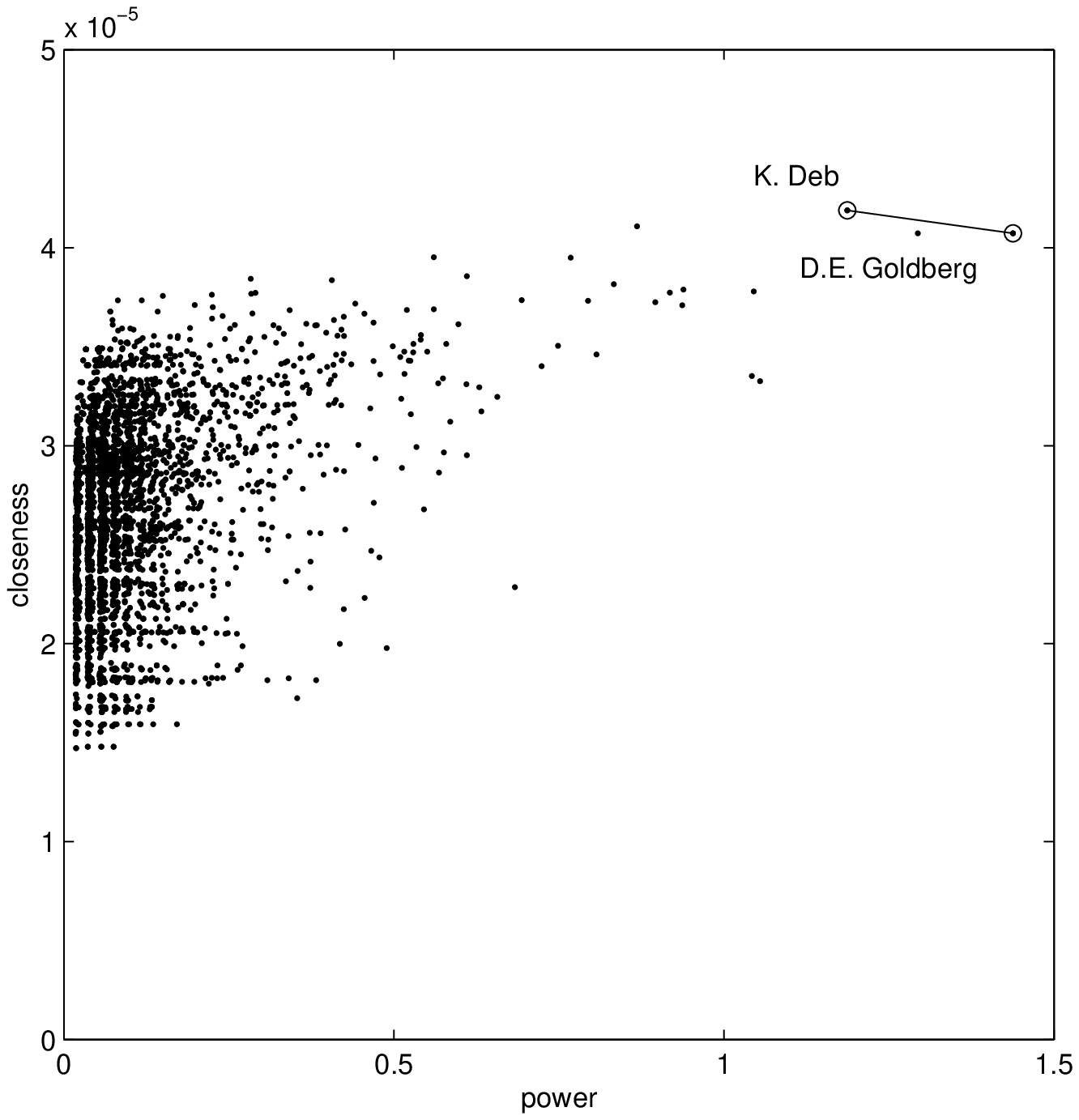}
\caption[]{Scatter plot of centrality values under different
measures. (Left) closeness vs. betweenness, (right) closeness vs.
power.\label{fig:pareto1}\vspace{1cm}}
\end{figure}

\begin{figure}[htbp]
\centering
\includegraphics[width=85mm]{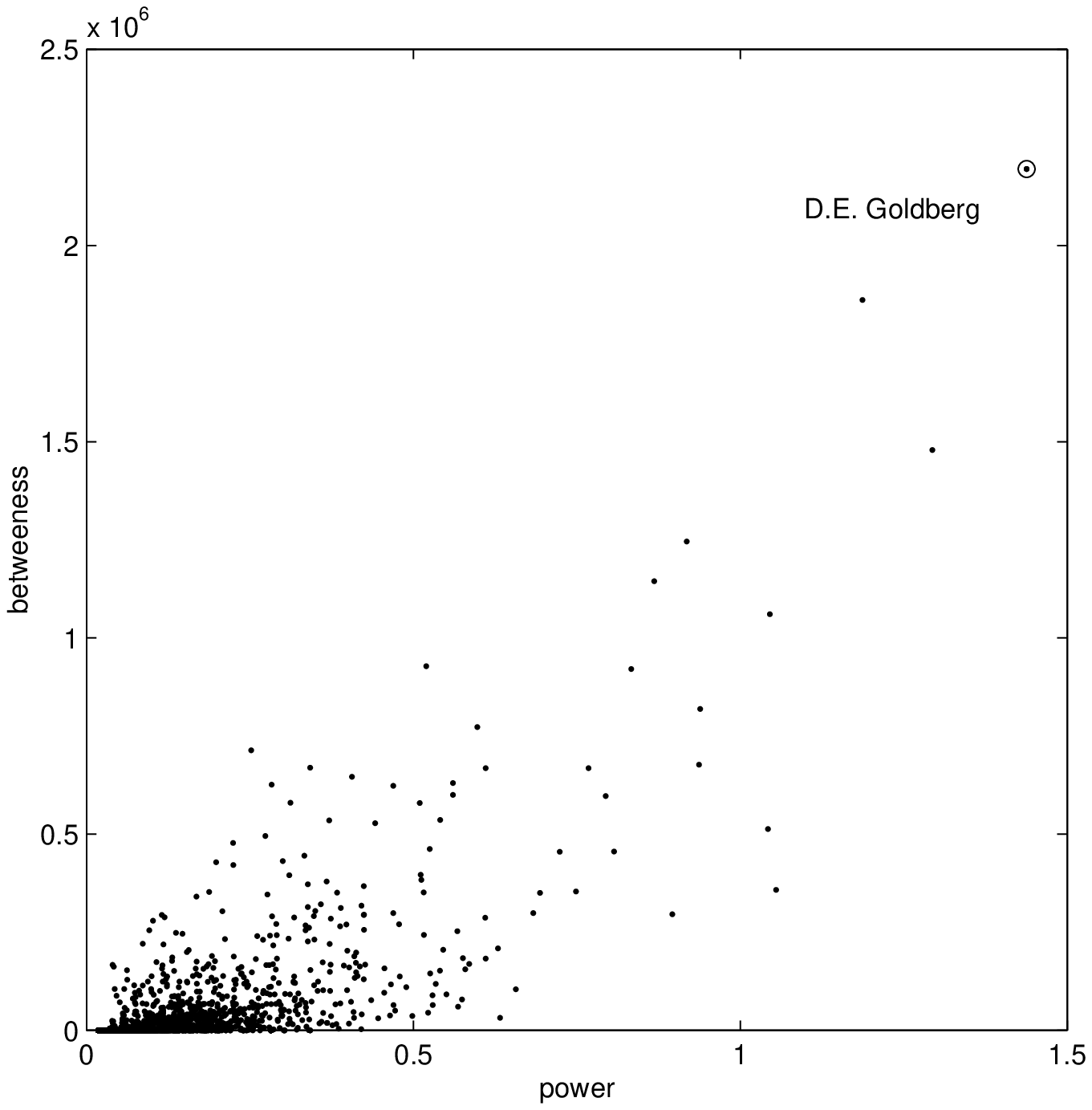}
\includegraphics[width=85mm]{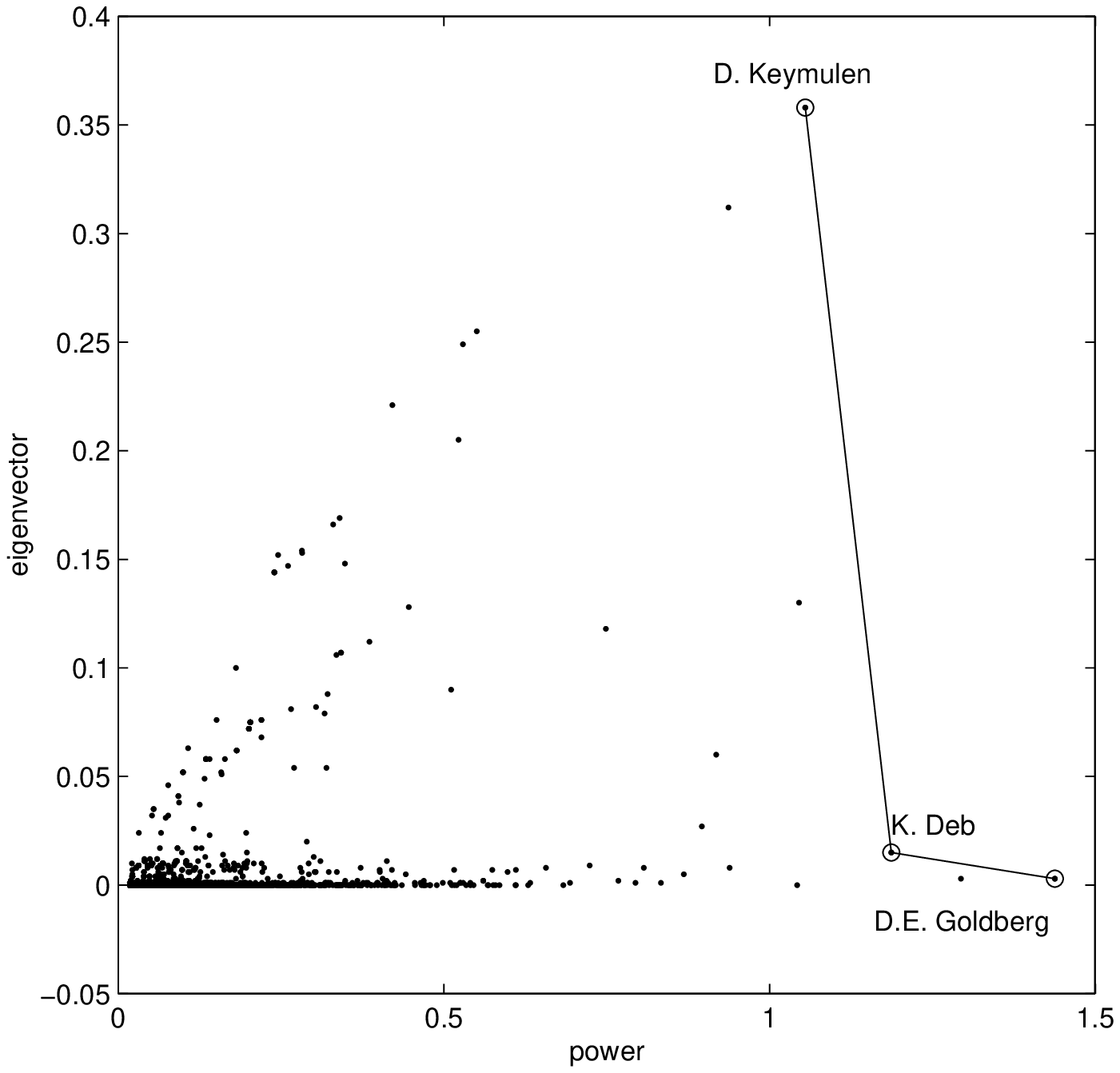}
\caption[]{Scatter plot of centrality values under different
measures. (Left) power vs. betweenness, (right) power vs.
eigenvector.\label{fig:pareto2}}
\end{figure}
\begin{figure}[htbp]
\centering
\includegraphics[width=85mm]{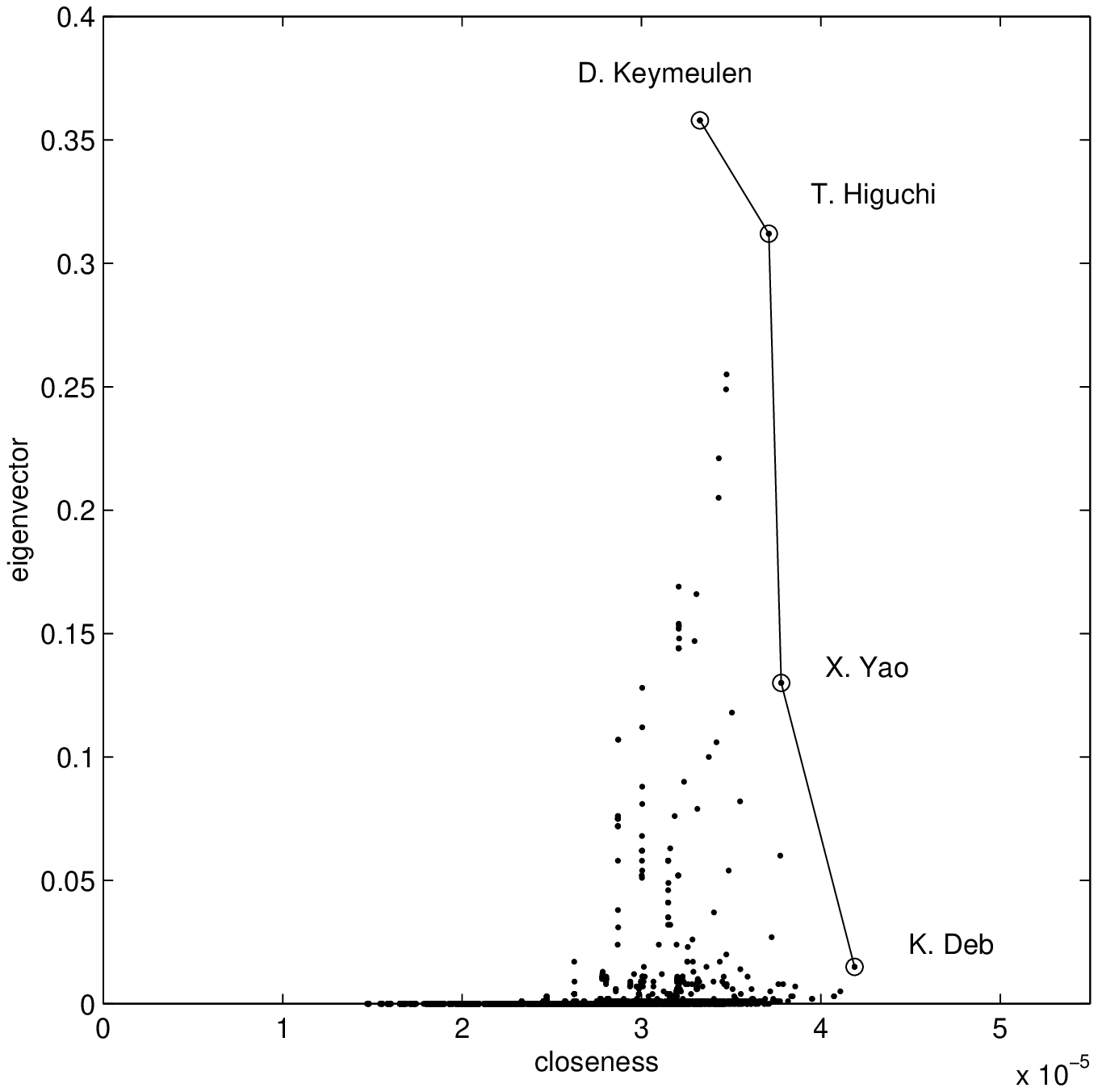}
\includegraphics[width=85mm]{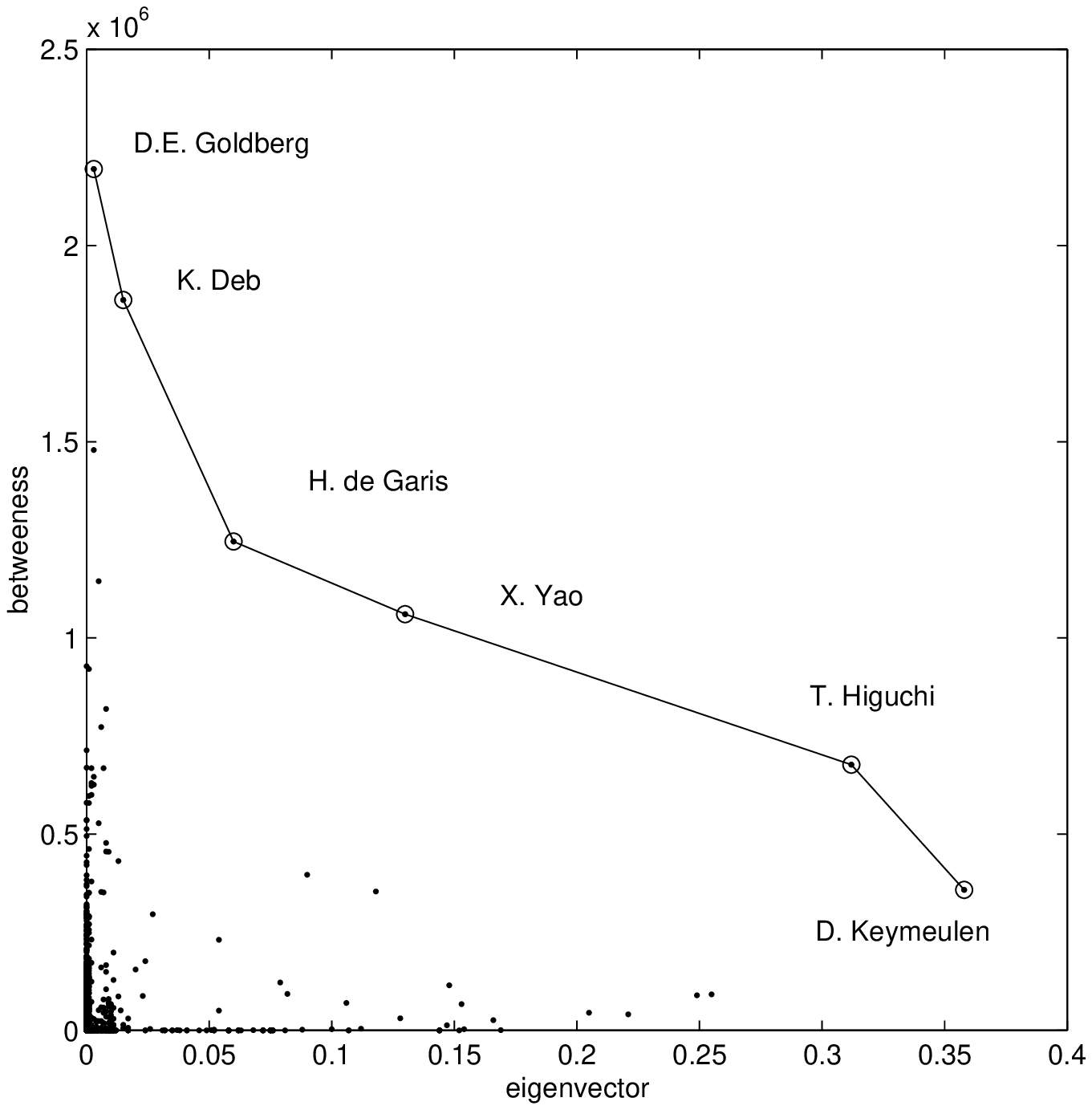}
\caption[]{Scatter plot of centrality values under different
measures. (Left) closeness vs. eigenvector (right), eigenvector vs.
betweenness.\label{fig:pareto3}}
\end{figure}

First of all, we have performed pairwise combinations of the
previous centrality measures, to determine the extent to which they
can be said to be correlated. Figures \ref{fig:pareto1},
\ref{fig:pareto2}, and \ref{fig:pareto3} show scatter plots of all
six possible combinations, and indicate the corresponding
non-dominated front (notice that centrality values are always
maximized). Notice firstly in Figure \ref{fig:pareto1} and Figure
\ref{fig:pareto2} (left) that Bonacich power, closeness, and
betweenness are fairly well correlated. The tail of the rankings is
rather wide, but a clear correlation is observed for top actors.
Indeed, there are just two actors appearing in the non-dominated
fronts: D.E. Goldberg (thrice) and K. Deb (twice). The former is
actually the global dominating actor for power vs. betweenness. When
eigenvector centrality is introduced, the situation is different:
centrality gravitates for this measure around a different set of
actors, and this is specifically clear in Figure \ref{fig:pareto2}
(right) and Figure \ref{fig:pareto3}. The correlation is now much
more questionable, and as a consequence the non-dominated fronts
tend to be wider.

Next, we consider the non-dominated front arising from the
simultaneous optimization of all four centrality measures. In this
case, the resulting actors are K. Deb, H. de Garis, D.E. Goldberg,
T. Higuchi, D. Keymeulen, and X. Yao. If we go to the second
non-dominated front (i.e., the front that would result from the
removal of actors in the previous front), we get T. B\"ack, C.
Coello Coello, D.B. Fogel, T. Hoshino, H. Iba, M. Iwata, L. Kang, J.
Li, Y. Liu, Z. Michalewicz, M. Schoenauer, and R. Salem. We can
easily identify highly recognized researchers in this list, as well
as some actors raised to a prominent status due to eigenvector
hitchhiking. To see the impact of this fourth centrality measure, we
have also recomputed the global fronts considering just closeness,
betweenness, and power. The result is the following:
\begin{itemize}
\item \textbf{Front \#1:} K. Deb, D.E. Goldberg
\item \textbf{Front \#2:} Z. Michalewicz, M. Schoenauer
\item \textbf{Front \#3:} T. B\"ack, A.E. Eiben, H. de Garis, D. Key-meulen, B.
Paechter, M. Tomassini, X. Yao
\item \textbf{Front \#4:} D.B. Fogel, J.J. Merelo, T. Higuchi, K.A. De Jong, L.
Kang, E. Lutton, R.E. Smith, L.D. Whitley
\item \textbf{Front \#5:} H.A. Abbass, H.-G. Beyer, J. Branke, M. Dorigo, T.C. Fogarty,
H. Iba, M. Keijzer, E.G. Talbi, M.D. Vose
\end{itemize}

Several things must be noted. Firstly, the top two actors for
eigenvector centrality also have prominent places in these fronts,
showing that the fact that they were spotted by the former
centrality measure is not a spurious effect, but a solid indicator
of their relevance in the network structure. Secondly, all
researchers appearing in these fronts are very well-known in the
field for their research excellence. Their appearance in one front
or another does not represent therefore a scientific ranking, but a
measure of their connectedness under three different measures.

\section{Conclusions}
\label{sec:conclusions} 

Centrality analysis is fundamental in order
to grasp the microscopic structure of a network, and understand the
mechanisms governing its temporal evolution. Consider that
our\footnote{Please note that the two authors are part of this community}
current network is the result of the incremental addition of ties
through the years. In this sense, two reflections can be made: the
first one is that central nodes owe their prominence to the fact
that they are strategically located within the network structure,
and this indicates that the growth of the network somehow gravitates
around them. Secondly, and related to the previous fact, centrality
is a dynamic property that changes with time. An asymmetric growth
of the network may displace one actor from a relatively central
position, or a highly active actor may become with time one of the
most relevant hubs of the network. The analysis presented in this
work must thus be interpreted as a snapshot of the situation.

There are two aspects to be highlighted in this work. The first one
refers to the methodology used. To the best of our knowledge, this
is the first time that a multi-objective approach has been used to
characterize centrality using different measures. We believe this is
the natural approach that should be followed in this kind of
studies, since each measure provides a different (despite some
obvious correlation) perspective on the relevance of each actor. The
second aspect is the actual results obtained. We have identified a
group of researchers (most prominently K. Deb and D.E. Goldberg, but
many others can be cited too), that systematically appear both in
the top rankings under different measures and in the corresponding
non-dominated fronts. This gives an objective picture of who the
best connected EC scientists were and are (recall that there is an
inherent historical component in centrality). This also gives hints
on the current ``hot'' points of activity, and indirectly (via an
examination of the corresponding research subareas) on the current
``hot'' topics. Related with these previous issues, and in
particular with the historical aspect of centrality, we have studied
elsewhere \cite{cotta07:gpem} the temporal evolution of the macroscopic properties of
the collaboration network. It would be very interesting, and it
actually constitutes one of our priorities, to conduct a temporal
analysis of centrality, identifying the trajectories of the most
relevant actors, and -if the trends are clear enough- even
forecasting future sociometric stars.

\section{Acknowledgments}

This paper has been funded in part by the Spanish MICYT projects  {\em
  NADEWeb: Nuevos Algoritmos Distribuidos Evolutivos en la web}, code 
TIC2003-09481-C04, NOHNES,  TIN2007-68083-C02-01, and P06-TIC-02025,
awarded by the regional government. 

\bibliographystyle{abbrv}
\bibliography{coauthorship}

\end{document}